\documentstyle[pre,aps,multicol]{revtex}
\input{epsf}
\begin{document}
\draft

\title{Quantum relaxation in open chaotic systems}

\author{Klaus M. Frahm}

\address {Laboratoire de Physique Quantique, UMR 5626 du CNRS, 
Universit\'e Paul Sabatier, F-31062 Toulouse Cedex 4, France}

\date{July 19, 1997; condmat/9707206, last revised October 5}

\maketitle

\begin{abstract}
Using the supersymmetry technique, we analytically derive the recent result 
of Casati, Maspero and Shepelyansky [cond-mat/9706103] 
according to which the quantum dynamics of open chaotic systems 
follows the classical decay up to a new quantum relaxation time 
scale $t_q\sim\sqrt{t_c\,t_H}$. This scale is larger than the classical 
escape time $t_c$ but still much smaller than the Heisenberg time $t_H$. 
For systems with orthogonal or unitary symmetry the quantum decay 
is slower than the classical one while for the symplectic case there 
is an intermediate regime in which the quantum decay is slightly faster. 
\end{abstract}


\begin{multicols}{2}
\narrowtext

The classical decay probability of a generic weakly open chaotic system 
obeys the exponential distribution $P_{\rm cl}(t)\propto e^{-t/t_c}$ 
where the mean escape time $t_c$ characterizes the effective coupling 
to the outside. Motivated by recent experiments on mesoscopic cavities 
or microwave billiards there has been renewed interest in the problem 
of {\em quantum life times} 
\cite{maspero,borgonovi,fyodorov1,fyodorov2,fyodorov3,piet2,mucciolo,comtet}. 
For example the quantum properties of ``chaotic'' maps with absorption 
\cite{maspero,borgonovi} were investigated. Recently, also 
analytical results for the statistical distributions of the complex 
poles of the scattering matrix \cite{fyodorov1,fyodorov3} or 
of the eigenvalues of the Wigner-Smith 
matrix of time delays \cite{piet2} were found for the 
generic problem of chaotic scattering. 

The problem of current relaxation in disordered metals 
which is similar to the quantum decay 
of an initially localized wave packet inside a chaotic cavity 
or a disordered region has been investigated by different 
analytical approaches \cite{altshuler,muzykantskii,mirlin}
in the framework of the 
nonlinear $\sigma$ model (replica or supersymmetry variant \cite{efetov}). 
In these works it was shown that the classical exponential decay is strongly 
supressed by quantum effects for time scales larger than the Heisenberg time 
$t_H$ giving rise to a log-normal distribution of relaxation times. 
Muzykantskii and Khmelnitskii \cite{muzykantskii} 
demonstrated that this is due to a 
nontrivial saddle point of the $\sigma$-model. For the case of open 
one-dimensional geometries they obtained for $t>t_H$ the behavior 
$P(t)\sim \exp[-g\ln^2(t/t_H)]$
where $g=t_H/t_c\gg 1$ is the conductance (in units of $e^2/h$). 
This result has an important relation to the probability to find 
a 'nearly localized state' in a normally metallic sample \cite{muzykantskii}. 
Also the quantum time evolution of open chaotic cavities was studied 
\cite{harney} giving a power law decay for 
$t\gg t_H/\min{\{T_j\}}$ where $0\le T_j\le 1$ are the 
transmission coefficients of the barrier by which the cavity is coupled 
to the outside. 

Recently, Casati, Maspero and Shepelyansky \cite{maspero} 
surprisingly found that for a quantum kicked rotator model 
with absorption \cite{borgonovi} significant deviations from the 
classical behavior already appear at an earlier {\em quantum relaxation 
time scale} $t_q\sim\sqrt{t_c\,t_H}\ll t_H$. 
Their argument \cite{maspero} is based on 
the complex eigenvalues of the non unitary time 
evolution operator being typically distributed in a narrow ring of width 
$E_c=1/t_c$ \cite{borgonovi} inside the unit circle. ($t_c\gg 1$ is measured 
in units of the kick period.) Then $t_q$ can be identified \cite{maspero} 
as the inverse of their typical distance in the complex plane. This picture 
is indeed supported by numerical quantum simulations \cite{maspero} for 
the kicked rotator. 

In this work, we present analytical results for a similar model 
by mapping it onto the supersymmetric nonlinear $\sigma$-model \cite{efetov} 
which is possible due to a recent progress of Altland and Zirnbauer 
\cite{zirn} for this type of systems. The $\sigma$-model in 
the zero-dimensional limit also 
applies to the case of a chaotic cavity coupled to external leads \cite{vwz}. 
We clearly confirm the findings of Casati et al. 
that the new time scale $t_q$ is indeed highly relevant for the problem 
and, additionally, we find at $t\sim t_q$ qualitatively 
different quantum effects 
for the three symmetry classes of random matrix theory \cite{mehta} 
which are characterized by 
the index $\beta=1$ for the orthogonal case (systems with time reversal 
symmetry and no spin mixing), $\beta=2$ for the unitary case (broken time 
reversal symmetry) and $\beta=4$ for the symplectic case (time reversal 
symmetry and strong spin mixing). 
This result supports the interpretation that the effect of 
weak localization (or anti-localization for $\beta=4$) can also 
be observed in open systems with absorption. 

We consider the quantum dynamics $|\psi(t+1)>\,= S |\psi(t)>$ of a 
generalized random phase kicked rotator model with the time evolution 
operator introduced in \cite{borgonovi,maspero}
\begin{equation}
\label{eq1}
S_{l\tilde l}=e^{i\mu_l}\,<l|\,e^{-i\,V(\theta)}\,|\tilde l>
\,e^{i\mu_{\tilde l}}
\end{equation}
where $l$, $\tilde l$ are the quantum numbers of ``angular momentum'' beeing 
conjugated to the angle $\theta$. As in Refs. \cite{borgonovi,maspero} the 
$l$-space is finite: $-N/2\le l,\tilde l\le N/2$ introducing  
effective absorption at the boundaries. We consider random 
phases $\mu_l$ 
\cite{footnote5} and a quite general periodic kick potential 
$V(\theta)$ (with a finite number of harmonics). 
The different symmetry classes are encoded in the 
symmetries of $V(\theta)$ \cite{footnote4}, 
i.e. $V(\theta)=V(-\theta)$ for 
$\beta=1$ and $V(\theta)=V_0(\theta)\openone_2
+\sum_{\nu=1}^3 \sigma_\nu\,V_\nu(\theta)$ 
for $\beta=4$. Here $\sigma_\nu$ are the Pauli 
matrices and $V_\nu(\theta)$ is even (or odd) for $\nu=0$ (or $\nu=1,2,3$). 
In the following, we consider the phase averaged quantity
$P(t)=\langle \left|<0|\,S^{\,t}\,|0>\right|^2\rangle_\mu$ 
to describe the decay of a quantum state initially localized at the site 
$l=0$. For short time scales this probability decays diffusively 
as $P(t)\propto 1/\sqrt{D\,t}$ (with diffusion constant 
$D=\langle V'(\theta)^2 \rangle_\theta\gg 1$) 
whereas for longer time scales and 
large system size ($t,N\gg D$) quantum localization leads to the 
saturation $P(t)\propto 1/\xi$ with the localization length 
$\xi=\beta D/2$. Here we concentrate on the case of a 
system size $N$ being much smaller than $\xi$ (i.e. 
$t_c\sim N^2/D\ll N=t_H$) and on time scales $t>t_c$. 

\begin{figure}
\epsfxsize=3.0in
\epsfysize=2.0in
\epsffile{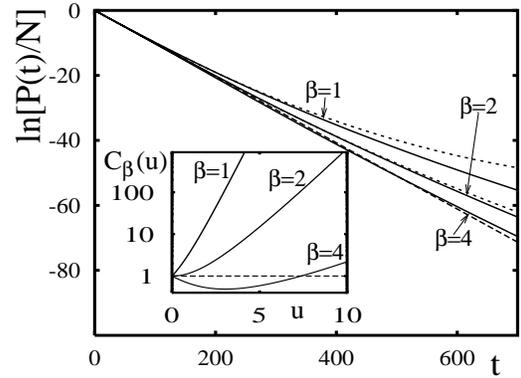}
\caption{Logarithm of $P(t)$ for the three symmetry classes 
with $E_{c,1} \approx 0.1$, $N=2000$ and all transmission eigenvalues 
$T_j=1$. The full lines are 
obtained from (\ref{eq16}) for $\beta=2$ or the corresponding 
integrals for $\beta=1,4$. The dashed line shows 
the classical exponential decay and the two dotted lines for $\beta=1,2$ 
correspond to Eq. (\ref{eq3}). For $\beta=4$, Eq. (\ref{eq3}) 
coincides with the full line. The inset shows the functions $C_\beta(u)$ 
given in Eqs. (\ref{eq4})-(\ref{eq6}). }
\label{fig1}
\end{figure}

We first present and discuss our main results before we outline 
some basic steps of the approach. We find 
that the first quantum corrections for $t_c< t\lesssim t_H^{2/3}\,t_c^{1/3}$ 
can be cast in the form 
\begin{equation}
\label{eq3}
P(t)\propto e^{-E_{c,1}\,t}\ C_\beta\left(\frac{E_{c,2}\,t^2}{2 N}
\right)\quad,\quad E_{c,\nu}=\frac{g_\nu}{N} 
\end{equation}
where for the 0d limit ($k\approx N/2$ or for a chaotic cavity) we have 
introduced the ``generalized conductance'' moments 
by $g_\nu=\sum_j\,T_j^{\,\nu}$, $\nu=1,2,\ldots$. 
Here the transmission eigenvalues $0\le T_j\le 1$ describe the effective 
coupling strength of the cavity with the boundary. For the 1d limit, 
$g_1=g_2=\pi^2 D/2N$ is (up to a numerical factor) the 
classical conductance from the site 0 to the boundary. 
The universal functions $C_\beta(u)$ have the form (inset 
of Fig. \ref{fig1})
\begin{eqnarray}
\nonumber
C_1(u) & = & \int_0^1 dx\int_0^1 dy\ \frac{2x(1-x)}{(1-x^2+x^2 y^2)^2}
\times\\
\label{eq4}
&& \times\exp\left[u\,\bigl(2x-(1-x^2+x^2 y^2)\bigr)\right]\\
& = & 1+u + {\textstyle \frac{5}{6}}\,u^2+\cdots,
\nonumber\\
\label{eq5}
C_2(u) & = & \sinh(u)/u= 1+{\textstyle \frac{1}{6}}\,u^2+\cdots,\\
\label{eq6}
C_4(u) & = & C_1(-u/2) = 1-
{\textstyle \frac{1}{2}}\,u + {\textstyle \frac{5}{24}}\,u^2+\cdots.
\end{eqnarray}
For $\beta=1,2$ the quantum probability $P(t)$ is above its classical 
value $P_{\rm cl}(t)$. The criterion $\ln[P(t_q)/P_{\rm cl}(t_q)]=0.1$ 
(see \cite{maspero}) to define the quantum relaxation 
time scale $t_q$ leads to 
$t_q\approx 0.45 \sqrt{N/E_{c,2}}$ for $\beta=1$ and 
$t_q\approx 1.24 \sqrt{N/E_{c,2}}$ for $\beta=2$. 
The numerical factor for $\beta=1$ is indeed close to $0.38$ found 
in \cite{maspero}. We note that for $\beta=2$ the function $C_2(u)$ has 
only a quadratic correction for small $u$. 
The situation for $\beta=4$ is particularly 
intriguing because here the quantum probability is 
initially even {\em below } the classical value. The function $C_4(u)$ has 
at $u_{\rm min}\approx 3.03$ ($t_{q,\rm min}\approx 2.46 \sqrt{N/E_{c,2}}$) 
its minimum value $0.488$ and 
it crosses the classical value $1$ again at $u_{\rm cr}\approx 7.36$ 
($t_{q,\rm cr}\approx 3.84 \sqrt{N/E_{c,2}}$). It seems that the linear 
term in the function $C_\beta(u)$ can be viewed as a weak-localization 
correction (anti-localization for $\beta=4$). 
In Fig. \ref{fig1}, we also show for the 0d case with $T_j=1$ 
the accurate result which is given by more complicated integrals
(see below). 

To derive these results, we have applied the supersymmetric technique 
\cite{efetov,vwz} which has recently been generalized \cite{zirn,klaus} 
to treat random phases instead of gaussian disorder. 
Repeating the steps described in Ref. \cite{klaus}, we can 
express the Laplace transform $\tilde P(\omega)$ of $P(t)$ as a 
functional integral of the type
\begin{equation}
\label{eq7}
\tilde P(\omega)=\int {\cal D}Q\ f\bigl(Q(0)\bigr)\ e^{-{\cal L}[Q]}.
\end{equation}
Here the integration is done over a field of $8\times 8$ supermatrices 
$Q(l)$, $-N/2\le l\le N/2$ with the non linear constraint 
$Q^2=1$ and particular symmetries for each universality 
class \cite{efetov}. $f\bigl(Q(0)\bigr)$ is a preexponential factor 
that depends only on the $Q$-field at site $0$. The action in (\ref{eq7}) 
has the form 
\begin{eqnarray}
\label{eq8}
{\cal L}[Q]&=&{\textstyle \frac{d}{2}}\ \mbox{Str}_{8N}\,
\ln\left(\hat B(\omega)+i\hat Q\right),\\
\label{eq9}
\hat B(\omega)&=&i\Lambda\ \frac{1-e^{i\omega/2}\,\hat U_0}
{1+e^{i\omega/2}\,\hat U_0}\quad,\quad \hat U_0=
\left(\begin{array}{cc}
U_0 &  \\
 & U_0^\dagger \\
\end{array}\right).
\end{eqnarray}
The number $d=1$ ($2$) for $\beta=1,2$ ($\beta=4$) measures the spin 
degeneracy and the supertrace extends over an $8N$-dimensional super space. 
$\hat Q$ is an operator containing the $Q(l)$-fields in its diagonal blocks 
and $U_0$ is a matrix with elements $<l|\,e^{-i\,V(\theta)}\,|\tilde l>
\otimes \openone_4$. 
The block structure in (\ref{eq9}) refers to the grading for advanced and 
retarded Greens functions with the matrix $\Lambda$ having the 
entries $+1$ ($-1$) in the upper (lower) diagonal block. 
As in \cite{klaus}, we expand the action in the limit of long wave lengths 
and long time scales which gives ${\cal L}[Q]\approx {\cal L}_B[Q]
+{\cal L}_{1d}[Q]$ 
where 
\begin{equation}
\label{eq10}
{\cal L}_{1d}[Q] = -\frac{d}{32}\int_{-\frac{N}{2}}^{\frac{N}{2}} dl
\,\mbox{Str}\Bigl(D(\partial_l Q)^2+4i\,\omega\,Q\Lambda\Bigr)
\end{equation}
is the standard one-dimensional $\sigma$ model action. 
Here the supertrace without subscript acts on $8\times 8$ supermatrices. 
The term ${\cal L}_B[Q]$ which was absent in \cite{zirn,klaus} 
arises from the boundary absorption because 
the operator $U_0$ is {\em not} unitary due to the cutoff in $l$ space. 
According to this we can write $\hat B(0)=B_1+i\Lambda B_2$ with 
hermitian matrices $B_1$ and $B_2$. Note that $B_2$ does not vanish because 
$U_0$ is not unitary. 
The boundary part of the action is then determined by the eigenvalues 
$0\le T_j^{(0)}\le 1$ of the hermitian matrix 
$\hat T^{(0)}=A^{-1/2}\ 4B_2\,A^{-1/2}$ (with $A=B_1^2+(1+B_2)^2$). 
These eigenvalues have the meaning of transparencies of 
coupling channels to the outside. Their precise distribution 
depends on microscopic details like system 
size and the particular choice of the kick potential $V(\theta)$. 
The eigenvectors with non vanishing $T^{(0)}_j$ have typically a 
support on the sites close to the boundary and the related boundary 
conductance 
$g^{(0)}=\sum_j T^{(0)}_j$ scales like the effective bandwidth of $U_0$: 
$g^{(0)}\sim \sqrt{D}$. We have verified this behavior by a numerical 
evaluation of $\hat T^{(0)}$ for the standard kicked rotator. 
Therefore we can write: 
${\cal L}_B[Q]=L_B(\hat T^{(0)},Q(\frac{N}{2}))+L_B(\hat T^{(0)},Q(-\frac{N}{2}))$ with  
\begin{equation}  
\label{eq12}  
L_B(\hat T^{(0)},Q)=\frac{d}{4}\sum_j \mbox{Str}\ \ln
(1+{\textstyle\frac{1}{2}}
\,T^{(0)}_j\ \Delta Q)
\end{equation}
and $\Delta Q={\textstyle \frac{1}{2}}(Q\Lambda+\Lambda Q)-1$. 
The sum runs over all non vanishing eigenvalues associated to one 
boundary. We note that for the $S$-matrix approach of Refs. \cite{vwz,iwz} 
exactly the same action is obtained where $T^{(0)}_j$ are the 
transmission eigenvalues of a tunnel barrier which 
couples a mesoscopic sample to an ideal quantum wire \cite{iwz}. 

The functional integral (\ref{eq7}) corresponds to a path integral which 
can be evaluated by solving a diffusion equation in $Q$-space
\cite{larkin,efetov,mmz}. Therefore we rewrite (\ref{eq7}) as 
\begin{equation}
\label{eq13}
\tilde P(\omega)=\int dQ\ f(Q)\ F^2(Q,N/2)
\end{equation}
where the function $F(Q,l)$ is determined 
by the partial differential equation \cite{larkin,efetov,mmz}
\begin{equation}
\label{eq14}
\partial_l F(Q,l)=\left(\frac{2}{\xi}\,\Delta_Q+
i\frac{d}{8}\,\omega\ \mbox{Str}(Q\Lambda)\right)\ F(Q,l)
\end{equation}
and the initial condition $F(Q,0)=\exp[-L_B(\hat T^{(0)},Q)]$. Here 
$\Delta_Q$ denotes the Laplace operator in $Q$-space (with the precise 
notations of Ref. \cite{piet1}). The general solution of (\ref{eq14}) for 
arbitrary frequencies is an involved mathematical problem. First, we 
consider the solution $F_0(Q,l)$ for the case $\omega=0$. 
For this, we note that $\exp[-L_B(\hat T,Q)]$ 
as a function of $T_j$ and $Q$ 
exactly coincides with the generating function (2.3) of Ref. \cite{piet1} 
which was used to prove the equivalence of the $\sigma$ model 
\cite{larkin,efetov,mmz} and Fokker-Planck approach 
\cite{dorokhov,mello,been_rev} for quasi one-dimensional 
disordered wires. According to the argumentation presented in \cite{piet1}, 
$F_0(Q,l)$ is exactly given by
\begin{equation}
\label{eq15}
F_0(Q,l)=\int d\hat T\ p(\hat T,l)\,\exp[-L_B(\hat T,Q)]
\end{equation}
where $p(\hat T,l)$ is a probability distribution of transmission 
eigenvalues $T_j$ which fulfills a certain Fokker-Planck equation 
(known as DMPK-equation due to Dorokhov \cite{dorokhov}, and 
Mello, Pereyra, Kumar \cite{mello}) 
with the initial condition $p(\hat T,0)=\delta(\hat T-\hat T^{(0)})$. 
$p(\hat T,l)$ describes the statistical transport properties of 
a quasi one-dimensional disordered wire in series with a tunnel barrier 
with transparencies $T^{(0)}_j$. At 
first sight (\ref{eq15}) seems to be more complicated due to the increased 
number of integrations. However, in the metallic limit, we can 
expand (\ref{eq12}) in powers of $\Delta Q$ with the self averaging 
transmission moments $g_1, g_2, g_3, \ldots$ as prefactors. 
Their ``quantum'' fluctuations are of order unity and have only 
an effect for $t\gtrsim t_H$. Therefore we can replace $g_\nu$ by their 
average values and omit the $T$-average.
These $g_\nu$-averages are in the classical 
limit determined by a set of differential equations which can be derived 
from the DMPK-equation \cite{been_rev}. To determine $F(Q,l)$ for 
$\omega\neq 0$ we use the expression for $F_0(Q,l)$ as an ansatz where the 
$g_\nu$ are now parameters to be determined as a function of $\omega$. 
The $\omega$-term only modifies the equation for $g_1$ giving: 
$g_1'(l)=-(2/D)\,g_1^2-i\omega$ and $g_2'(l)=(4/D)(g_1^2-2\,g_1\,g_2)$. 
Omiting the details, we mention that the explicit solutions determine 
$F(Q,l)$ and thus provide a closed expression for 
$\tilde P(\omega)$ as one $Q$-integral (\ref{eq13}). Using the 
standard parameterizations 
for $Q$ introduced by Efetov \cite{efetov}, we can express (\ref{eq13}) as 
an integral over two ($\beta=2$) or three ($\beta=1,4$) radial parameters. 
We can perform the integrations for $\omega$ (from the Fourier transform) 
and for the 
effective variable $s=\mbox{Str}(\Delta Q)$ in a saddle point approximation 
which is justified for $t\gg t_c$. Keeping the first two terms 
with $g_1$ and $g_2$ in $F(Q,l)$ we obtain our main result 
(\ref{eq3}-\ref{eq6}) for the 1d case. The situation for the 0d case 
is much easier, here we can simply insert the given 'boundary' 
transmission eigenvalues and perform the $\omega$-integration. 
For lack of space, we only state the result for $\beta=2$ 
\begin{eqnarray}
\label{eq16}
P(t) & \approx & \frac{1}{t}\int_0^{\min(1,\frac{t}{N})}
dx\ \left(1+2\frac{t}{N}-2x\right)\ e^{-L(x)},\\
\label{eq17}
L(x) & = & \sum_j\ \ln\left(\frac{1+(\frac{t}{\tilde N}-x)\,T_j}
{1-x\,T_j}\right).
\end{eqnarray}
The corresponding expressions for $\beta=1,4$ 
have a similar structure with two integrations. 
The curves shown in Fig. \ref{fig1} were obtained from a numerical 
evaluation of these integrals. They also lead to our principal 
result (\ref{eq3})-(\ref{eq6}) 
if we expand the logarithm in (\ref{eq17}) up to second order in $T$. 
The expansion parameter 
here is in principle $t/\tilde N\sim t/t_H\ll 1$. However, one can estimate 
that the third order term gives a contribution $\propto t^3/(t_c\,t_H^2)$ 
which has to be smaller than unity because of the exponential in (\ref{eq16}).
Of course the same criterion holds for the 1d case if we restrict ourselves 
to the first two moments $g_1$ and $g_2$. 

In summary, we have found that for open chaotic systems the first quantum 
corrections to the classical relaxation 
process appear at a quantum relaxation time scale 
$t_q\sim \sqrt{t_c\,t_H}$ with different 
effects for each universality class (Fig. \ref{fig1}). 
This scale is determined by the second moment of 
transmissions eigenvalues $T_j$ describing the effective 
coupling strength of the initial site with the boundary. It would be 
very interesting to relate this finding more clearly to the physical mechanism 
suggested in Ref. \cite{maspero} according to which $t_q$ is the time 
scale at which the quantum discreteness of the complex eigenvalues 
$\exp(iE_j-\Gamma_j/2)$ of the non unitary time evolution operator $S$ 
\cite{borgonovi} can be resolved. We emphasize 
that in view of the universal $\sigma$ model formulation 
our results apply not only to the kicked rotator model (\ref{eq1}) but 
also to chaotic cavities (corresponding to the zero-dimensional random 
matrix limit) and to quasi one-dimensional disordered wires. 
In this case one should consider the time evolution of 
a wave packet of plane waves in an 
energy interval of size $\hbar/\tau$ where $\tau$ is the elastic scattering 
time. The typical extension of a such a wave packet is just 
the mean free path which is in any case the smallest length scale 
that can be resolved by the standard $\sigma$ model \cite{efetov}. 

Due to the almost identical $\sigma$ model action it is important 
to understand the relation of our results with the approach of 
Ref. \cite{muzykantskii} where mainly the limit $t>t_H$ was 
considered. A recent careful analysis \cite{private} 
of the saddle point approach pioneered in Ref. \cite{muzykantskii} 
indeed gives for the regime $t_q\ll t\ll t_H$ the 
behavior $\ln P(t)\approx -(t/t_c)\,[1- t/(\beta\,g\,t_c)]$ confirming 
Eqs. (\ref{eq3})-(\ref{eq6}) for $u\gg 1$. 
Furthermore, for $t>t_H$ we can state that the log-normal behavior found in 
\cite{altshuler,muzykantskii,mirlin} 
should also apply to the {\em average} decay rate for the kicked 
rotator model. However, for very long time 
scales one should also focus on the distribution of the decay function 
because for a {\em given} sample the decay 
is then again exponential with a decay rate given by the minimal 
$\Gamma_j$ \cite{maspero}. 

Concerning the zero dimensional limit, (\ref{eq3}),(\ref{eq4}) for 
$\beta=1$ is in principle also contained in the 
exact integral expressions of \cite{harney}. However, since the 
corresponding limit was not worked out there the time scale $t_q$ 
remained undetected. We emphasize that here the $T_j$ are given model 
parameters and $E_{c,2}$ might parameterically be smaller than $E_{c,1}$ if 
all $T_j\ll 1$. We mention that very recently Savin and Sokolov 
\cite{savin} independently also found the time scale $t_q$ in the frame work 
of the supersymmetric approach. 
Their results which apply for the 0d case with unitary symmetry 
completely agree with our findings (\ref{eq3}) and (\ref{eq16}). 

The author acknowledges D.~L.~Shepelyansky and B.~Georgeot for fruitful 
and inspiring discussions.

\end{multicols}


\begin{thebibliography}{99}
\bibitem{maspero} G. Casati, G. Maspero, and D. L. Shepelyansky, 
	preprint cond-mat/9706103.

\bibitem{borgonovi} F. Borgonovi, I. Guarneri and D. L. Shepelyansky, 
	Phys. Rev. A {\bf 43}, 4517 (1991). 

\bibitem{fyodorov1} Y. V. Fyodorov and H.-J. Sommers, J. Math. 
	Phys. {\bf 38}, 1918 (1997). 

\bibitem{fyodorov2} Y. V. Fyodorov and H.-J. Sommers, Phys. Rev. Lett. 
        {\bf 76}, 4709 (1996); Y. V. Fyodorov, D. V. Savin, and H.-J. 
	Sommers, Phys. Rev. E {\bf 55}, 4857 (1997). 

\bibitem{fyodorov3} Y. V. Fyodorov, B. A. Khoruzhenko, and H.-J. Sommers, 
	Phys.Lett. A {\bf 226}, 46 (1997). 

\bibitem{piet2} P. W. Brouwer, K. M. Frahm, and C. W. J. Beenakker, 
	Phys. Rev. Lett. {\bf 78}, 4737 (1997). 

\bibitem{mucciolo} E. R. Mucciolo, R. A. Jalabert, and J.-L. Pichard, 
	preprint condmat/9703178.

\bibitem{comtet} A. Comtet and C. Texier, preprint condmat/9707046.

\bibitem{altshuler} B. L. Altshuler, V. E. Kravtsov, and I. V. Lerner, 
	JETP Lett. {\bf 45}, 199 (1987); Sov. Phys. JETP {\bf 67}, 
	795 (1988); I. E. Smolyarenko and B. L. Altshuler, Phys. Rev. 
	B {\bf 55}, 10451 (1997). 

\bibitem{muzykantskii} B. A. Muzykantskii, and D. E. Khmelnitskii, Phys. Rev. 
	B {\bf 51}, 5481 (1995); preprint cond-mat/9601045.

\bibitem{mirlin} A. D. Mirlin, Pis'ma Zh. Eksp. Teor. Fiz. 
	{\bf 62}, 583 (1995).

\bibitem{efetov} K. B. Efetov, Adv. in Phys. {\bf 32}, 53
	(1983); {\it Supersymmetry in Disorder and Chaos}, 
	Cambridge University Press (1997).

\bibitem{harney} H. L. Harney, F.-M. Dittes, and A. M\"uller, Ann. Phys. (NY) 
	{\bf 220}, 159 (1992). 

\bibitem{zirn} A. Altland and M. R. Zirnbauer, Phys. Rev. Lett. 
	{\bf 77}, 4536 (1996); M. R. Zirnbauer, J. Phys. A {\bf 29}, 
	7113 (1996). 

\bibitem{vwz} J. M. Verbaarschot, H. A. Weidenm\"uller, and 
	M. R. Zirnbauer, Phys. Rep. {\bf 129}, 367 (1985).

\bibitem{mehta} M.L. Mehta, {\it Random Matrices} 
        (Academic Press, New York, 1991).

\bibitem{footnote5} In this way complications related to a finite chaos 
	border or stable islands in phase space are avoided. 

\bibitem{footnote4} Note that for the quantum kicked rotator the 
	angle $\theta$ corresponds to the quasi momentum of $l$. 

\bibitem{klaus} K. M. Frahm, Phys. Rev. B {\bf 55}, R8626 (1997). 

\bibitem{iwz} S. Iida, H. A. Weidenm\"uller, and J. A. Zuk, Ann.
	Phys. (NY) {\bf 200}, 219 (1990).

\bibitem{larkin} K. B. Efetov and A. I. Larkin, Zh. Eksp. Theor. Fiz. {\bf 85},
	764 (1983) [Sov. Phys. JETP {\bf 58}, 444 (1983)].

\bibitem{mmz} A. D. Mirlin, A. M\"uller-Groeling, and
        M. R. Zirnbauer, Ann. Phys. (NY) {\bf 236}, 325 (1994).

\bibitem{piet1} P. W. Brouwer and K. M. Frahm, Phys. Rev. B {\bf 53}, 
	1490 (1996). 

\bibitem{dorokhov} O. N. Dorokhov, JETP Lett. {\bf 36}, 318 (1982).

\bibitem{mello} P. A. Mello, P. Pereyra, and N. Kumar, Ann.
	Phys. (NY) {\bf 181}, 290 (1988).

\bibitem{been_rev} C. W. J. Beenakker, Rev. Mod. Phys. {\bf 69}, 731 (1997). 

\bibitem{private} A. D. Mirlin, B. A. Muzykantskii, and D. E. Khmelnitskii, 
	private communication. 

\bibitem{savin} D. V. Savin and V. V. Sokolow, preprint cond-mat/9707187, 
	to appear in Phys. Rev. E, Rap. Comm.

\end{thebibliography}
\end{document}